\documentstyle[aps,psfig,amsmath,amssymb,times]{revtex}
\begin{document}
\title{Symmetry breaking and turbulence in perturbed plane Couette flow}
\author{Laurette S. Tuckerman}
\address{Laboratoire d'Informatique pour la M\'ecanique et les Sciences de
l'Ing\'enieur (LIMSI-CNRS)\\ Orsay France (laurette@limsi.fr)}
\author{Dwight Barkley}
\address{Mathematics Institute, University of Warwick, Coventry CV4 7AL, 
United Kingdom \\(barkley@maths.warwick.ac.uk)}
\maketitle
\begin{abstract}
Perturbed plane Couette flow containing a thin spanwise-oriented
ribbon undergoes a subcritical bifurcation at $Re \approx 230$ to a steady
3D state containing streamwise vortices.  This bifurcation is followed
by several others giving rise to a fascinating series of stable and
unstable steady states of different symmetries and wavelengths.  
First, the backwards-bifurcating branch reverses
direction and becomes stable near $Re \approx 200$. Then, the spanwise
reflection symmetry is broken, leading to two asymmetric branches
which are themselves destabilized at $Re \approx 420$.  Above this 
Reynolds number, time evolution leads first to a metastable state 
whose spanwise wavelength is halved and then to complicated 
time-dependent behavior. These features are in agreement with experiments.  
\end{abstract}
%
%
\section{Introduction}

Research on plane Couette flow has long been hampered by the absence
of states intermediate in complexity between laminar plane Couette
flow and three-dimensional turbulence.  Intermediate states can be
created, however, if a thin wire oriented in the spanwise direction is
inserted in an experimental setup \cite{Bottin97,Bottin98}.  No
longer subject to Squire's theorem, this perturbed configuration
undergoes a bifurcation to a three-dimensional steady or quasi-steady
state.  The 3D states contain vortices oriented in the streamwise
direction and of finite streamwise extent, localized around the
ribbon.  As the wire radius is reduced, the Reynolds number threshold
for the bifurcation and the streamwise extent occupied by the vortices
increase, while the range of Reynolds numbers over which the 3D steady
states exist decreases.  

We have carried out a numerical study corresponding to the experiments
of \cite{Bottin97,Bottin98}, focusing primarily on the largest wire
radius used. In a previous study \cite{Barkley}, we carried out a
linear and weakly nonlinear stability analysis of this configuration.
Here, we present a detailed bifurcation diagram for this case.  Our
calculations show a rich bifurcation structure with many types of
solutions: stable and unstable; steady, periodic, and aperiodic; and
of different symmetries.  Some of these solutions persist as the wire
radius is reduced.

\section{Methods}

The time-dependent Navier-Stokes equations have been solved using the
spectral element code {\tt Prism} \cite{Henderson} written by
Henderson.  Instead of a wire, we have used a ribbon of infinitesimal
streamwise ($x$) extent whose cross-channel ($y$) height is taken
equal to the diameter of the wire and whose length in the homogeneous
transverse ($z$) direction is infinite.  
For most of the current study the computational domain is $\vert x
\vert \leq 32$, $\vert y \vert \leq 1$, and $\vert z \vert \leq \lambda/2$
where $\lambda=2\pi/1.3$ is approximately the numerically determined
critical wavelength.
Periodic boundary conditions have been imposed at
$x=\pm 32$ and at $z=\pm \lambda/2$ and no-slip conditions at the
channel walls $y=\pm 1$ and at the ribbon $x=0$, $|y|\leq \rho$.
The ratio $\rho$ of the ribbon height to that of the channel is set 
to $\rho = 0.086$ except where otherwise specified.  In
the $(x,y)$ directions, we use $24 \times 5$ computational elements, 
each of which
is covered by a grid of $7 \times 7$ collocation points or
interpolating polynomials.  In the $z$ direction, we use 32 Fourier
modes or gridpoints. (Simulations were also conducted in a reduced
spanwise domain $\vert z \vert < \lambda/4$ with 16 Fourier modes.)
This leads to a total of 143840 gridpoints or basis functions per
velocity component.  
To compute each asymptotic state required between 500 and 10000
nondimensional time units (i.e. units of channel width/velocity
difference), which in turn required between 3 and 60 CPU
hours on each of 16 processors of a Cray T3E.
Some tests of the adequacy of our numerical resolution and streamwise
domain size are reported in \cite{Barkley}.  The resolution for
complex three-dimensional flows has been checked by increasing the
polynomial order from 7 to 11 and the number of Fourier modes from 32
to 64.

\section{Bifurcation scenario for $\rho = 0.086$}

\begin{figure}
\vspace*{-1cm}
\centerline{
\psfig{file=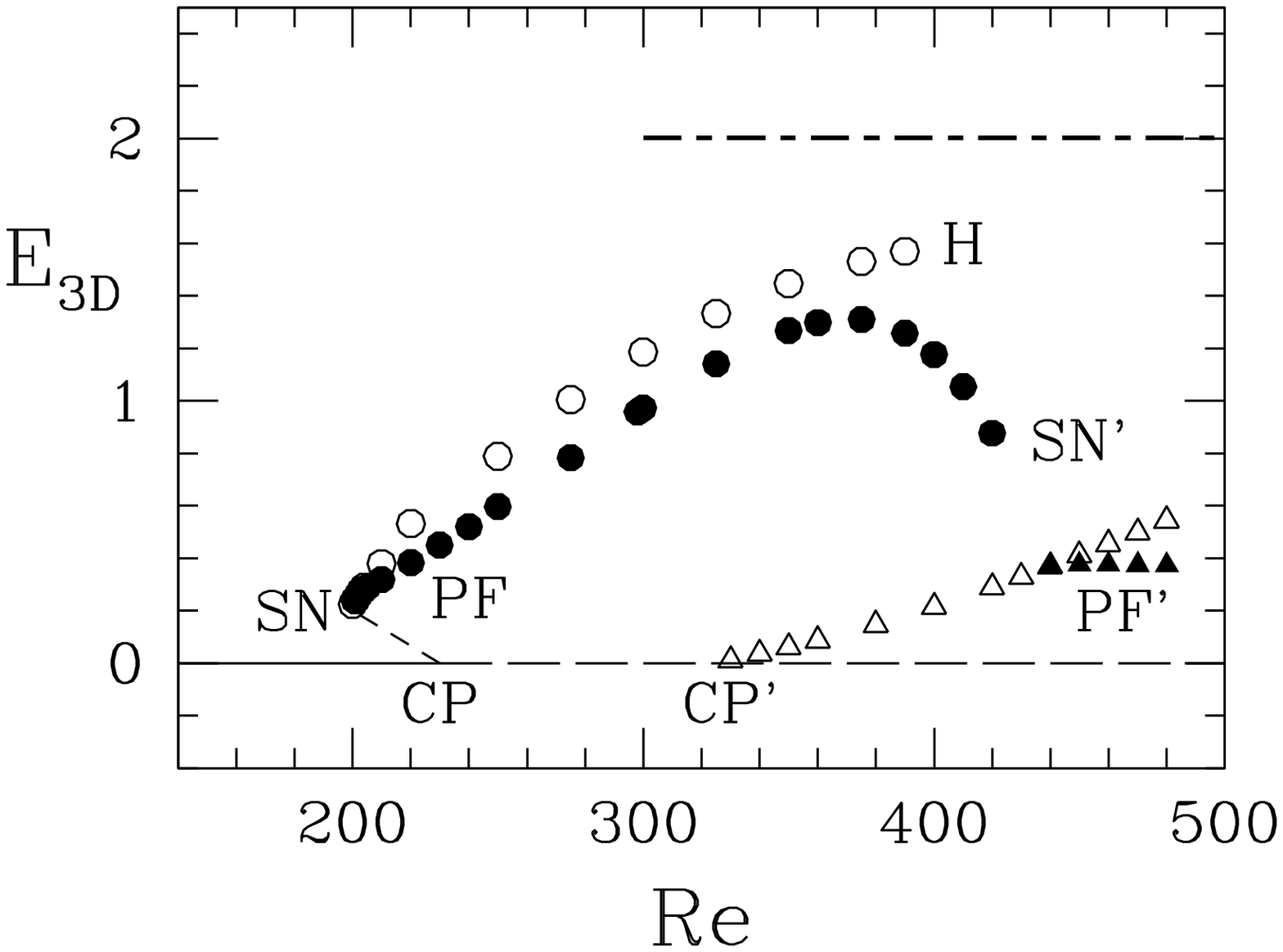,width=8cm}
}
\caption{ Bifurcation diagram for perturbed plane Couette flow with
ribbon height $\rho=0.086$. \\ 
$\bullet$Horizontal line: 2D states
exist for all $Re$; they are stable for $Re<Re_{\rm CP}=228$. \\
$\bullet$Short-dashed line (schematic): unstable symmetric states with
spanwise wavelength $\lambda$ exist between $Re_{\rm SN}=197$ and
$Re_{\rm CP}=228$. \\ 
$\bullet$Hollow circles: symmetric states with
$\lambda$ exist between $Re_{\rm SN}=197$ and $Re_{\rm H} = 395$;
they are stable between $Re_{\rm SN}$ and $Re_{\rm PF} = 201$. \\
$\bullet$Filled circles: asymmetric states with $\lambda$ exist
between $Re_{\rm PF} = 201$ and $Re_{\rm SN'} = 420$. \\
$\bullet$Hollow triangles: symmetrict states with $\lambda/2$ exist
for $Re > Re_{\rm CP'}=330$. \\ 
$\bullet$Filled triangles: asymmetric
states with $\lambda/2$ exist for $Re > Re_{\rm PF'} =440$. \\
$\bullet$Heavy long-and-short-dashed line (schematic): turbulent
states occur for $Re > Re_{\rm CH} = 300$.  }
\label{combined}
\end{figure}

Figure \ref{combined} shows $E_{3D}$, the energy in the $z$-dependent
modes for all the steady states we have calculated for $Re<500$, and
serves as a bifurcation diagram.  Each branch is distinguished by its
symmetry.  The geometry and basic 2D flow have O(2) symmetry in the
periodic spanwise direction $z$, i.e. translations $z\rightarrow
z+z_0$ and reflections $z\rightarrow -z$, In the $(x,y)$ plane, they
have centrosymmetry $(x,y)\rightarrow(-x,-y)$.

The 2D branch loses stability via a circle pitchfork bifurcation at
$Re_{\rm CP}=228$, breaking the translation symmetry in $z$.  The
critical spanwise wavelength is approximately $\lambda=4.8$ 
and the critical wavenumber $\beta = 1.3$.  
The circle pitchfork bifurcation is subcritical, and so the 3D states
created branch leftwards and are unstable; we cannot calculate them
with the methods used here.  These states have reflection symmetry in
$z$ and centro-symmetry in $(x,y)$; we call them 3D symmetric states.
The centro-symmetry can visualized as follows: At the ribbon location
at $x=0$, four small vortices are present. The upper two vortices
persist for $x>0$, while the lower two persist for $x<0$.  

\begin{figure*}[ht]
\vspace*{-7cm}
\centerline{
\psfig{file=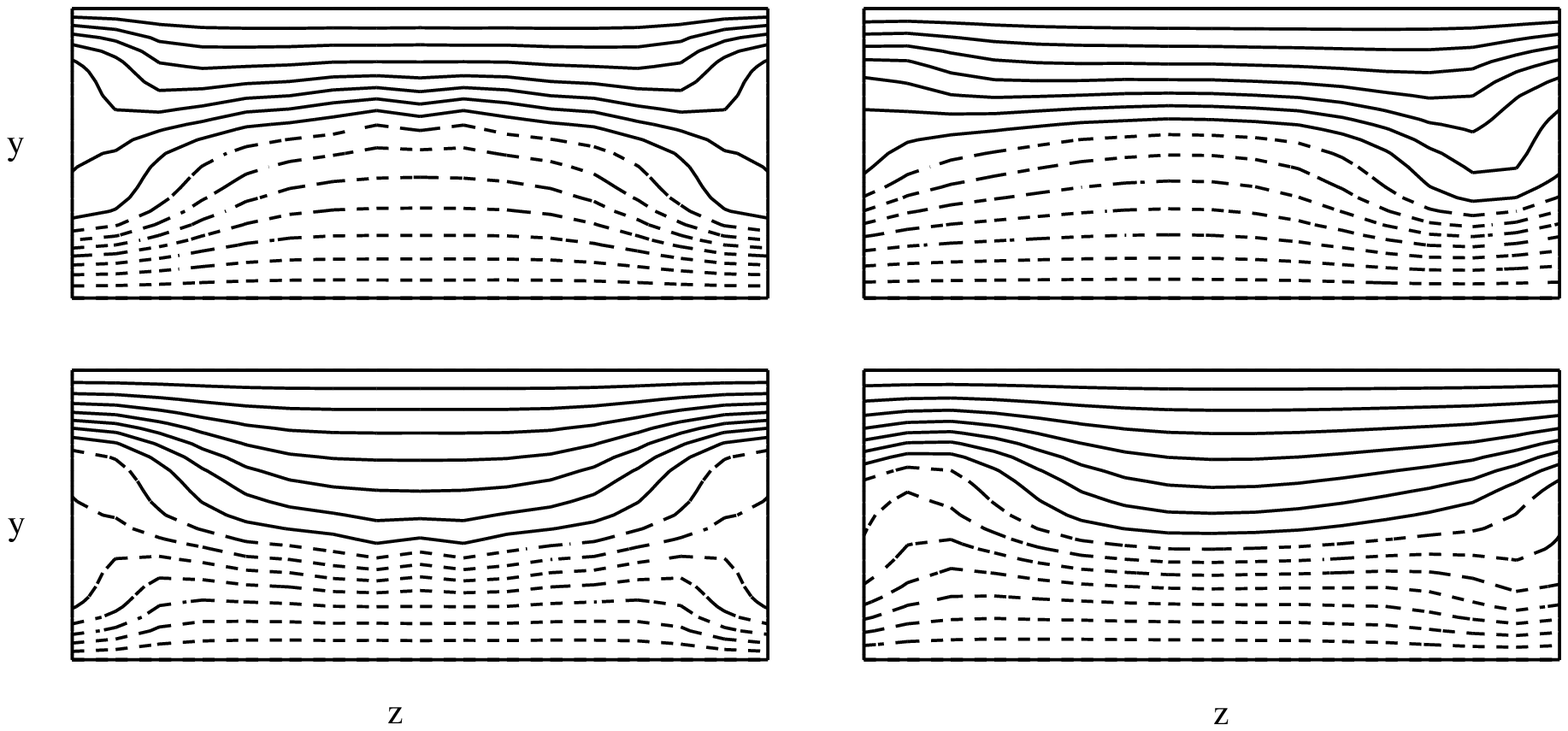,width=16cm}}
\vspace*{-6cm}
\caption{ Two velocity fields at $Re=240$, illustrating breaking of
reflection symmetry in $z$.  Contours of streamwise velocity $u$ are
shown at $x=2$ (above) and at $x=-2$ (below).  Left: State with
reflection symmetry in $z$ and centro-symmetry in $(x,y)$.  For $x=2$,
deformation of $u$ contours shows that $w$ velocity is upwards at
mid-$z$. Thus vortex on left (right) is counter-clockwise (clockwise).
For $x=-2$, the direction of $w$ and vortex orientation are
reversed. \\ Right: State with only the pointwise symmetry
$(x,y,z)\rightarrow (-x,-y,-z)$.  }
\label{240}
\vspace*{-7cm}
\centerline{
\psfig{file=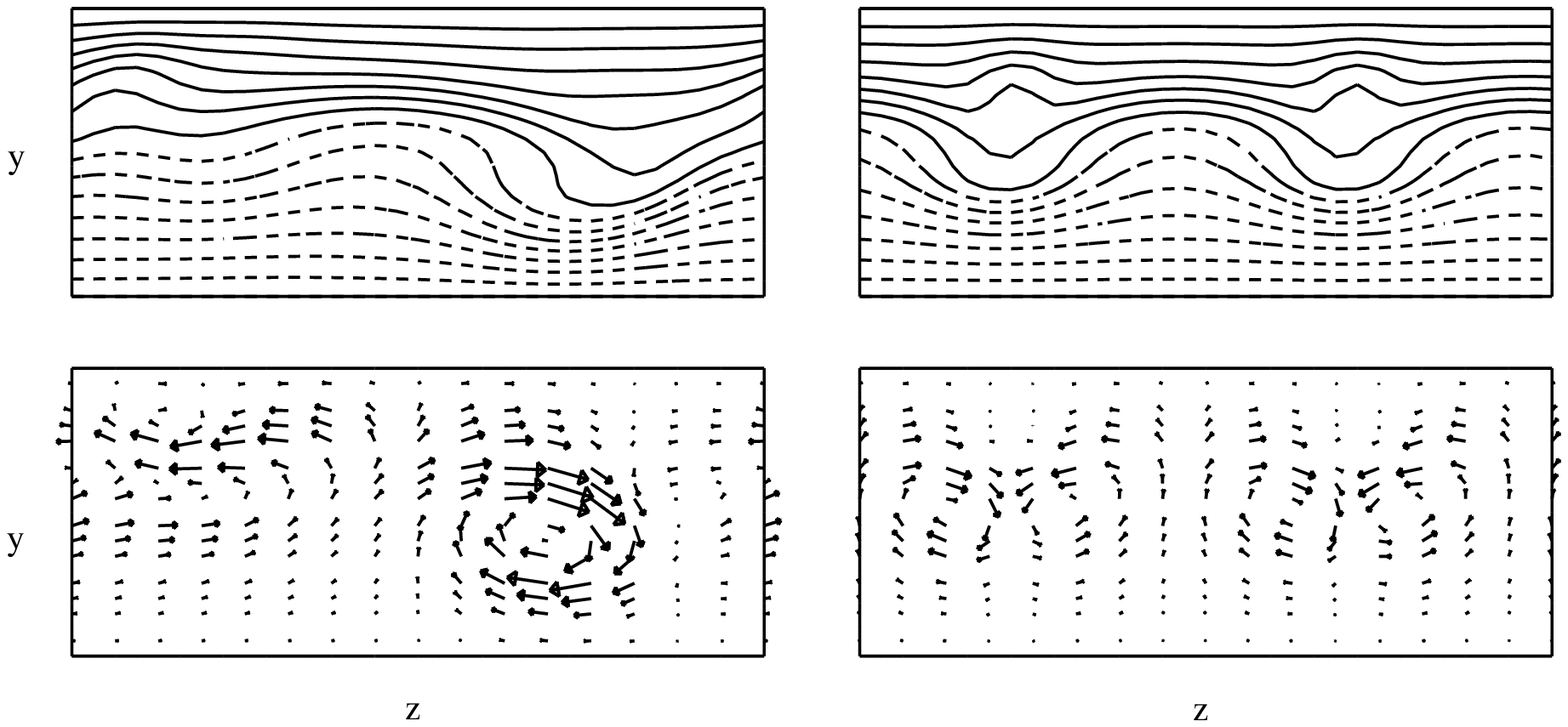,width=16cm}}
\vspace*{-6cm}
\caption{ Velocity fields at $Re=400$ and $Re=450$ illustrating
symmetry-restoring transition.  Contours of streamwise velocity $u$
are shown above and $(v,w)$ velocity field vectors are shown below,
both at $x=2$.  Left: Asymmetric state with spanwise wavelength
$\lambda=4.8$ at $Re=400$. The asymmetry in $z$ is very
pronounced. \\ Right: Metastable state with spanwise wavelength
$\lambda/2=2.4$ at $Re=450$.  The vortices occupy only the central
portion in $y$.}
\label{400-450}
\end{figure*}

\begin{figure*}[ht]
\vspace*{-6cm}
\centerline{
\psfig{file=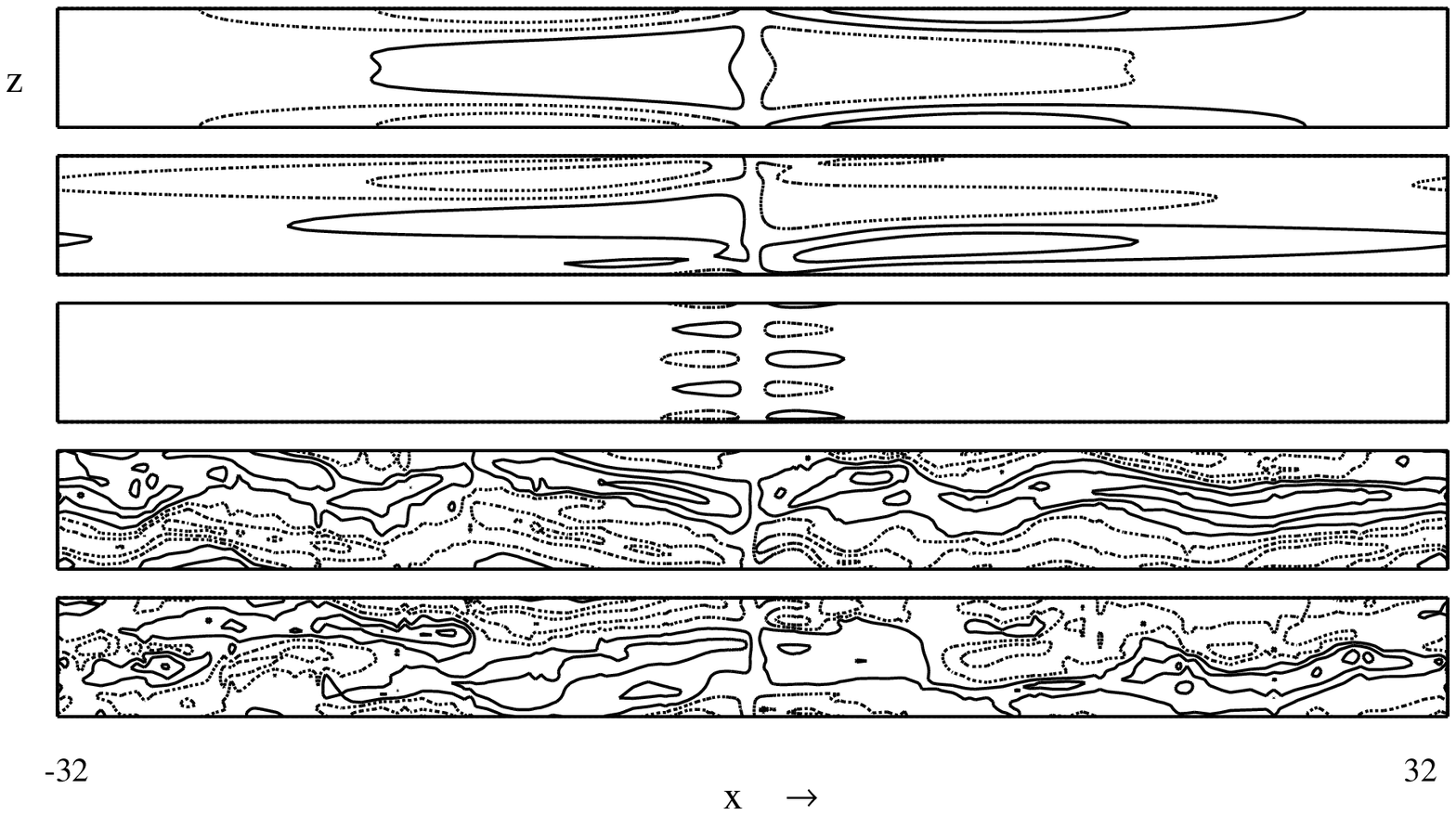,width=16cm}}
\vspace*{-6cm}
\caption{
Streamwise velocity contours at $Re=350$ in the $y=0$ midplane.
Shown from top to bottom are the symmetric, the asymmetric, and 
the $\lambda/2$ steady states, instantaneous turbulent
fields at $t=355$ and $t=630$ where $E_1$ is locally maximal and
minimal, respectively.
}
\label{R350y}
\vspace*{-8cm}
\centerline{
\psfig{file=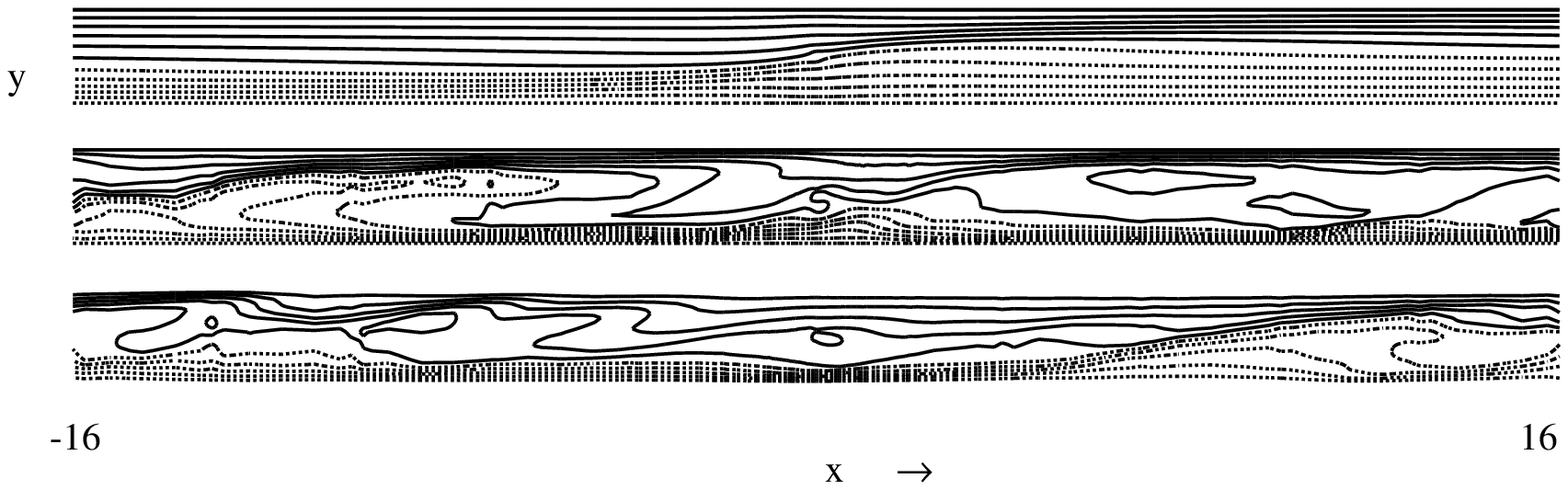,width=18cm}}
\vspace*{-7cm}
\caption{
Streamwise velocity contours at $Re=350$ in the $z=0$ plane
in the central half of the domain.
Shown from top to bottom are the asymmetric steady state and 
instantaneous turbulent
fields at $t=355$ and $t=630$ where $E_1$ is maximal and
minimal, respectively.
}
\label{R350z}
\end{figure*}

The 3D branch changes direction and is stabilized by a saddle-node
bifurcation at $Re_{\rm SN}=197$.  Its stability is short-lived,
however, lasting only until a pitchfork bifurcation at $Re_{\rm
PF}=201$.  The pitchfork bifurcation creates new stable branches with
only the pointwise symmetry $(x,y,z)\rightarrow(-x,-y,-z)$; we call
these 3D asymmetric states.  Figure \ref{240} illustrates this
symmetry breaking by showing two different velocity fields at
$Re=240$.  The symmetric 3D field on the left has two different
reflection symmetries, satisfying both $u(x,y,-z)=u(x,y,z)$ and
$u(-x,-y,z)=-u(x,y,z)$.  The asymmetric 3D field on the right
satisfies only the single reflection symmetry $u(-x,-y,-z)=-u(x,y,z)$.
The difference between symmetric and asymmetric 3D fields can also be
seen in figure~\ref{R350y} discussed below.

Although the symmetric 3D branch is unstable, we can continue to
calculate it by imposing reflection symmetry in $z$.  It is further
destabilized, however, by a Hopf bifurcation at $Re_{\rm H}\approx
395$, beyond which we have not followed it.

\begin{figure}[ht]
\vspace*{-1cm}
\centerline{
\psfig{file=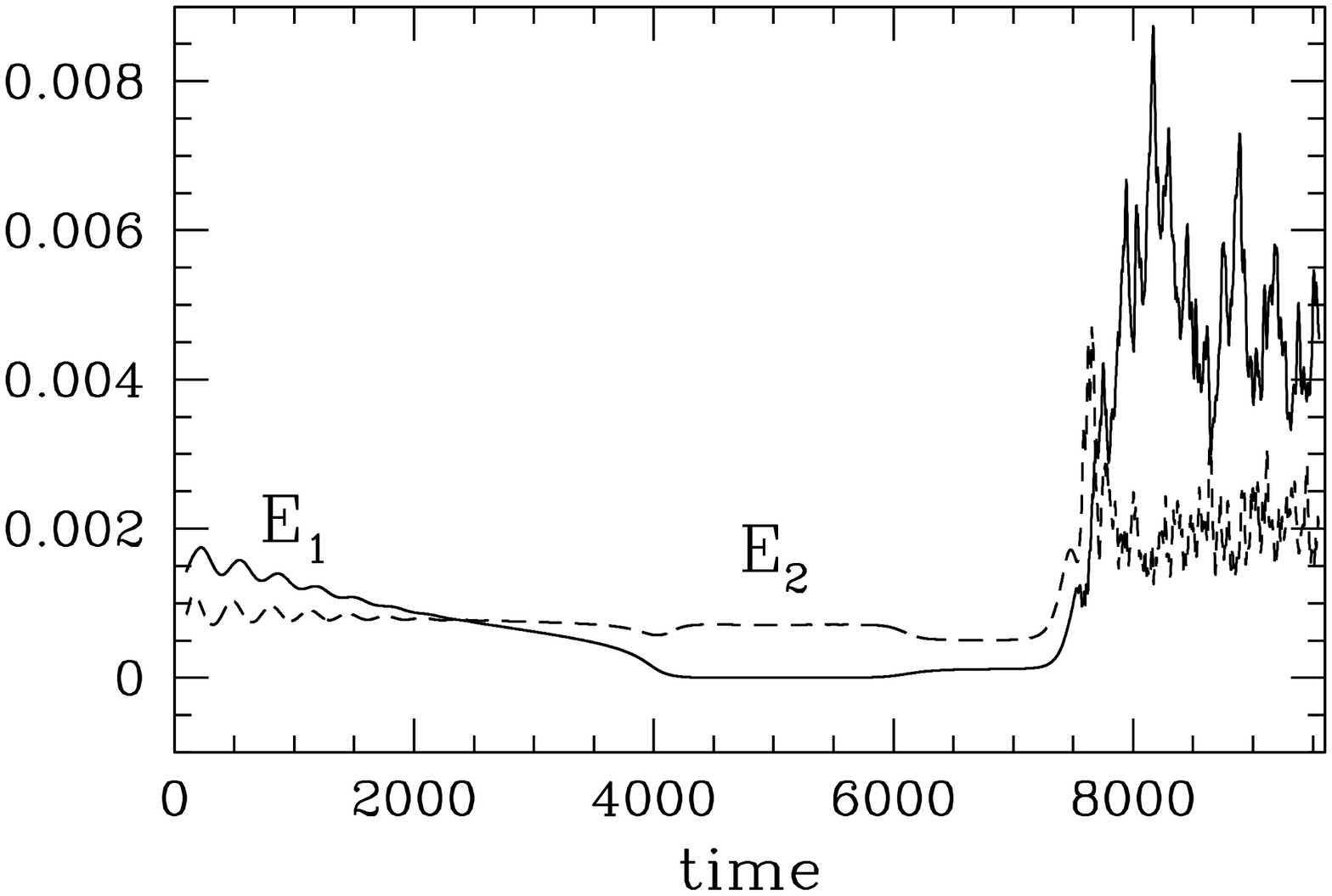,width=8cm}}
\vspace*{-.5cm}
\caption{ Time evolution of energy in the spanwise Fourier components
$\beta$ ($E_1$, solid curve) and $2\beta$ ($E_2$, dashed curve) at
$Re = 450$.  Initial state is asymmetric $\beta$ state at $Re =
400$.  $E_1$ decreases to near-zero levels for $4300 \lesssim t
\lesssim 5800$; during this interval, the flow is in a metastable
state with wavenumber $2\beta$.  For $t \gtrsim 7500$, the flow
undergoes large-amplitude irregular oscillations, corresponding to
turbulence.  }
\label{timeseries}
\centerline{
\psfig{file=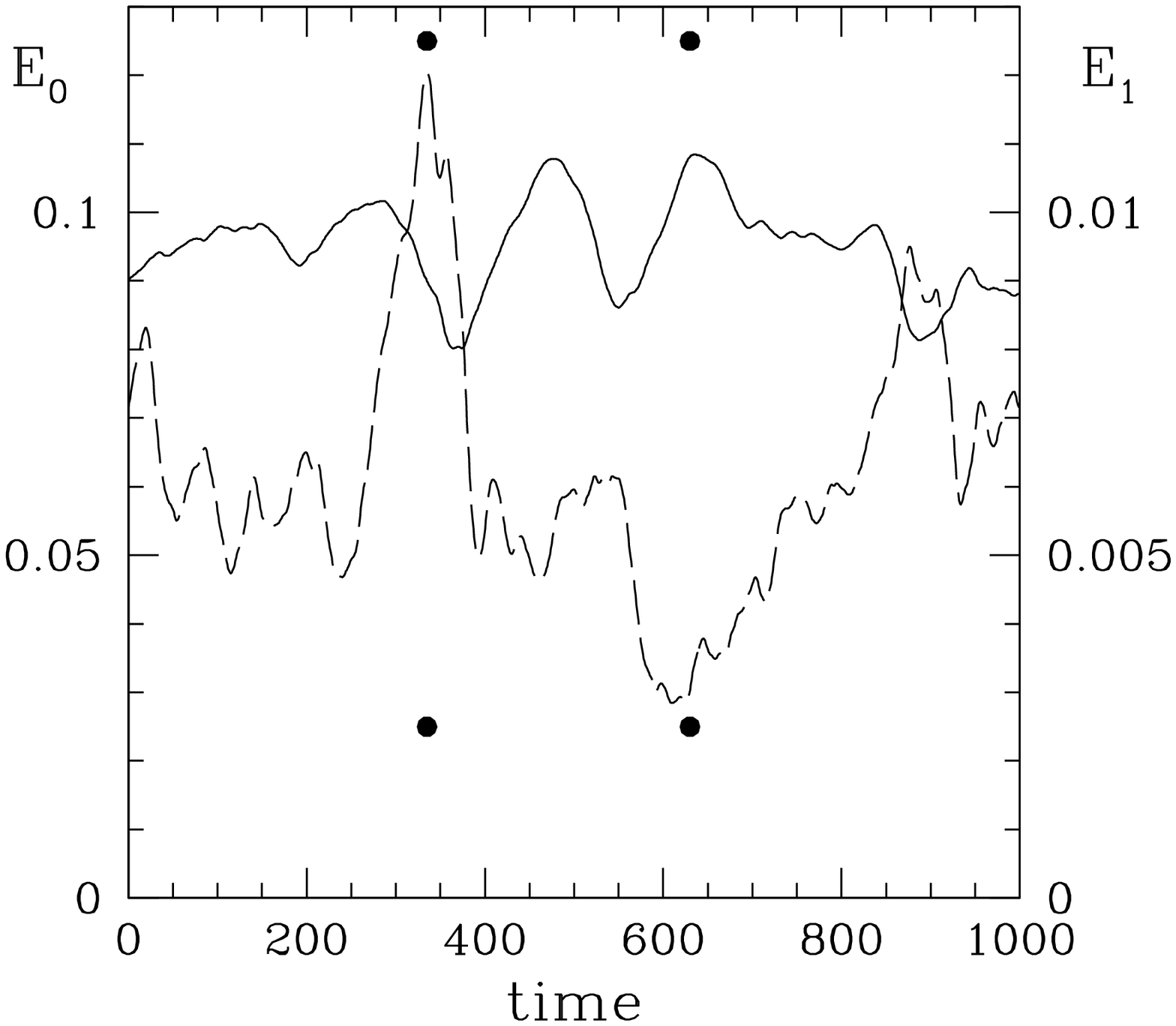,width=8cm}}
\caption{
Evolution of energy $E_0$ (solid curve) and $E_1$ (dashed curve)
in the zero and $\beta$ spanwise Fourier modes for a turbulent 
state at $Re=350$. Dots at $t=355$ and $t=630$ show times at
which instantaneous streamwise velocity contours are plotted
in figures \ref{R350y} and \ref{R350z}.
}
\label{events}
\end{figure}

The asymmetric 3D branches change direction and are destabilized by a
second saddle-node bifurcation at $Re_{\rm SN'}\approx 420$.
Surprisingly, time-dependent simulation at $Re=450$ from an initial
asymmetric state at $Re=400$ leads to a metastable state with half the
imposed wavelength of $\lambda=4.8$, or equivalently, twice the
wavenumber of $\beta=1.3$.  The evolution in time of the energies
$E_1$ and $E_2$ in the $\beta$ and $2 \beta$ spanwise Fourier
components is shown in figure \ref{timeseries}.  The initial field at
$Re=400$ and the metastable state at $Re=450$ are shown in figure
\ref{400-450}.  This transition is {\it symmetry-restoring} since the
field is invariant under translation in $z$ by $\lambda/2$.  The
metastable state persists during $4300 \lesssim t \lesssim 5800$, when
$E_1$ is near zero.

The metastable state is the $\lambda/2$ branch created from the 2D
branch by a circle pitchfork bifurcation at $Re_{\rm CP'}=330$; see
figure \ref{combined}.  Calculations show that it branches rightwards.
Each of the two halves of the field is symmetric under reflection in
$z$ about its midplane $z=\lambda/4$ or $z=-\lambda/4$.  The
$\lambda/2$ branch undergoes another pitchfork bifurcation at $Re_{\rm
PF'}=440$, analogous to that undergone at $Re_{\rm PF}$, creating
branches which do not have this reflection symmetry.  From figure
\ref{400-450}, it can be seen that the vortices in the $\lambda/2$
field remain somewhat circular; their cross-channel height is reduced
along with their spanwise extent.  The absence of vortices near the
upper and lower walls could indicate that the streamwise velocity
profile is more stable in this region than that near the center.

When the $\lambda/2$ branch is created at $Re_{\rm CP'}$, it is
necessarily unstable to wavelength doubling in a domain of size
$\lambda$, because its parent 2D branch is already unstable to
$\lambda$ modes.  We are able to calculate the $\lambda/2$ branch from
its creation at $Re_{\rm CP'}$ (and, for $Re > Re_{\rm PF'}$, its
asymmetric version) by using a domain of size $\lambda/2$.  However,
we emphasize that during its appearance as a metastable state, it is
calculated in the full domain of size $\lambda$, i.e.  it has been
stabilized to wavelength doubling.
We propose a possible mechanism for this stabilization: At $Re_{SN'}$,
the asymmetric $\lambda$ branches change direction and stability.  If
the unstable asymmetric $\lambda$ branches terminate on the symmetric
$\lambda/2$ branch in a subcritical pitchfork bifurcation $Re_{PF''}$,
then the symmetric $\lambda/2$ branch will be stabilized to $\lambda$
perturbations for $Re> Re_{\rm PF''}$, as will the asymmetric
$\lambda/2$ branch for $Re > Re_{\rm PF'}$.

However, this bifurcation scenario sheds no light on the subsequent
evolution from the metastable $\lambda/2$ state to irregular
oscillations, as shown in figure \ref{timeseries} for $t \gtrsim
7500$.  Irregular oscillations persist when the Reynolds number is
reduced until $Re < Re_{CH}=300$, where the flow reverts to the steady
asymmetric $\lambda$ branch.  We believe these states to correspond
essentially to the turbulent flow observed in unperturbed plane
Couette flow for $Re \gtrsim 325$
\cite{Lundbladh,Tillmark,Daviaud,Hamilton}, both because of the
closeness of the lower bound in Reynolds number, and because of their
appearance and streamwise extent (see figure~\ref{R350y}).  Because
lack of resolution can produce spurious time dependence, we have
verified that these dynamics persist with increased numerical
resolution.

Figure \ref{events} shows the evolution of the energy $E_0$ and $E_1$
in the zero and $\beta$ spanwise Fourier components at $Re=350$.
Turbulent states at $Re=350$ are illustrated via streamwise velocity
contours in figures \ref{R350y} and \ref{R350z}.  Figure \ref{R350y}
shows the symmetric, asymmetric, and $\lambda/2$ steady states, in the
$(x,z)$ midplane ($y=0$), as well as two instantaneous snapshots of
turbulent states $t=355$ and $t=630$, where $E_1$ is locally maximal
and minimal, respectively (see figure \ref{events}).  Figure
\ref{R350z} shows the asymmetric steady state and two instantaneous
snapshots of turbulent states at $t=355$, 630 in the $(x,y)$ midplane
($z=0$).  The deviation from plane Couette flow is highly
localized around the wire at $x=0$ for the steady states, but 
extends over the entire streamwise domain for the turbulent states.

In the experiments \cite{Bottin97,Bottin98}, streamwise vortices are
observed for $Re>150$, compared to our threshold of $Re=200$.  Their
wavelengths are between 5.2 and 5.7, as compared with our critical
$\lambda \approx 4.8$.  Intermittency is observed experimentally for
$Re>280$ and turbulence for $Re>325$, in close agreement with our
observation of $Re>300$.  Spatial period-halving events are also
observed in the experiments \cite{private}, as in our transition from
the $\lambda$ to the $\lambda/2$ metastable state. This period-halving
transition should be amenable to bifurcation-theoretic analysis.

\section{Smaller ribbons}

An important question is the dependence of this scenario on ribbon
height $\rho$. Results from calculations performed for two smaller
heights, $\rho = 0.071$ and $\rho = 0.043$, are given in Table
\ref{rhodep}.  For both of these ribbon heights, the 2D flow undergoes
a circle pitchfork bifurcation to a 3D eigenmode.  Because the
critical wavenumber, denoted by $\beta_c$, changes very little with
$\rho$, we have given more precise values for $\beta_c (\rho)$ in
Table \ref{rhodep} in order to specify the $\rho$ dependence.  The
fact that $\beta_c$ is relatively insensitive to $\rho$ provides
evidence for the idea that this instability arises from the underlying
unperturbed plane Couette flow rather than depending sensitively on
the perturbing ribbon.  (We have used the same domain with spanwise
periodicity $\lambda=2\pi/(1.30)$ to calculate all secondary
bifurcations.)  The critical Reynolds number $Re_{\rm CP}$ increases
as $\rho$ decreases, as expected from the absence of linear
instability at finite $Re$ for plane Couette flow. Both aspects of the
$\rho$-dependence of the transition to three dimensionality are also
observed in the experiments.

For $\rho = 0.071$, the bifurcation scenario is similar to that at
$\rho = 0.086$.  $Re_{\rm SN}$ increases more slowly as $\rho$
decreases than does $Re_{\rm CP}$, lending support to the hypothesis that
branches of steady 3D states might continue to exist as $\rho$
approaches zero, although disconnected from the 2D branch.  However,
$Re_{SN'}$ decreases, so that the Reynolds number range over which
{\it stable} steady 3D states exist is smaller, as is also observed
experimentally.  Specifically, the branch of unstable 3D states
bifurcating subcritically from the 2D flow occupies the Reynolds
number range $Re_{\rm CP}-Re_{\rm SN}$, which increases from 31 for
$\rho = 0.086$ to 65 for $\rho = 0.071$, while the branches of stable
3D states occupy the Reynolds number range $Re_{\rm SN'}-Re_{\rm SN}$
which decreases from 223 for $\rho = 0.086$ to 147 for $\rho = 0.071$.
We note that \cite{Nagata90,Nagata98} and \cite{Eckhardt} have
attempted to compute steady states of plane Couette flow containing
streamwise vortices by continuing Taylor vortex flow.  In the plane
Couette flow limit, they find that the solutions which persist are
analogues of wavy Taylor vortex flow, here streamwise traveling waves,
and exist for $Re>125$.

For $\rho=0.043$, we have been unable to find any stable steady 3D
states, despite extensive searching.  Thus Table \ref{rhodep} lists
only the primary instability $Re_{\rm CP}$ and $\beta_c$ for $\rho =
0.043$.  We observe irregular oscillations for all the initial
conditions, Reynolds numbers, and spatial resolutions we have tried.
This is in contrast to the experiments \cite{Bottin97,Bottin98}, in
which approximately steady states containing streamwise vortices are
observed for $\rho$ even smaller than 0.043.  Calculations at
intermediate ribbon heights would clarify whether, when, and how the
stable steady 3D states disappear as $\rho$ is reduced.

\begin{table}
\begin{center}
\begin{tabular}{|c||c|c||c|c|c|c|}
$\rho$ & $\beta_c$ & $Re_{\rm CP}$ & $Re_{\rm SN}$ & $Re_{\rm PF}$ & $Re_{\rm SN'}$ & $Re_{\rm H}$ \\ \hline
0.086 & 1.28 & 228  & 197 & 202 & 420 & 395\\
0.071 & 1.30 & 283  & 218 & 218 & 365 & 395 \\
0.043 & 1.45 & 538  & & & &  
\end{tabular}
\end{center}
\caption{Dependence of bifurcations on ribbon height $\rho$.
See figure \ref{combined} and text for description of bifurcation sequence.
All secondary bifurcations (to the right of the second double vertical line) 
have been calculated in a domain of spanwise periodicity length 2$\pi$/(1.30).
}
\label{rhodep}
\end{table}

\section{Acknowledgements}
We gratefully acknowledge Ron Henderson for the use of {\tt Prism}
and IDRIS/CNRS (Institut du Developpement et des Ressources en Informatiques 
Scientifiques, Centre National de la Recherche Scientifique) 
for the computational facilities.

\end{document}